
\mag = 1200					
\vsize=9.6truein \hsize=6.3truein			


\overfullrule=0pt    				
\parskip = 2ex plus .5ex minus .1ex
\parindent=0pt



\fontdimen16\tensy=2.7pt
\fontdimen17\tensy=2.7pt


 at 29.9 truept			
 at 24.9 truept			
\font\vbig = cmr10 at 20.7 truept			
\font\bbbig = cmr10 at 17.3 truept			
\font\bbig =cmr10 at 14.4 truept		
\font\med = cmr10 at 11 truept	
\font\small = cmr10 at 10 truept	
 at 10 truept	
 at 11 truept	
 at 14.4 truept	

 at 11 truept
 at 14.4 truept
 at 20.7 truept
 at 24.9 truept
 at 29.9 truept
 at 10 truept
 at 14.4 truept
 at 14.4 truept
 at 17.3 truept
 at 10 truept


\font\tenbit=cmmib10
\font\eightbit=cmmib10 at 10truept
	\textfont10=\tenbit \scriptfont10=\eightbit
	
\newfam\bitfam  \textfont\bitfam\tenbit
	\scriptfont\bitfam\eightbit


\mathchardef\bsigma="A1B


\def\da{\downarrow}

\def\dg{\dagger}

\def\la{\langle}

\def\ra{\rangle}

\def\tm{\times}
\def\ua{\uparrow}


\def\al{\alpha}
\def\be{\beta}
\def\dl{\delta}

\def\ep{\epsilon}

\def\ga{\gamma}

\def\si{\sigma}


\def\fd{f^{\dagger}} 
\def\Pt{P_{\tau}}

\def\Ptp{P_{{\tau}^{\prime}}}
\def\Qtp{Q_{{\tau}^{\prime}}}

\def\Qtd{Q_{\tau}^{\dagger}}
\def\Ptpd{P_{{\tau}^{\prime}}^{\dagger}}
\def\Qtpd{Q_{{\tau}^{\prime}}^{\dagger}}
\def\Aot{A_{\tau}^0}
\def\Atv{{\bf A}_{\tau}}
\def\Bot{B_{\tau}^0}

\def\Aots{A_{\tau}^{0 \ast}}
\def\Bots{B_{\tau}^{0 \ast}}

\def\Btsv{{\bf B}_{\tau}^\ast}
\def\At{A_{\tau}}
\def\Bt{B_{\tau}}
\def\Ats{A_{\tau}^\ast}
\def\Bts{B_{\tau}^\ast}

\def\Atuv{{\bf {\hat A}}_{\tau}}
\def\Atuvs{{\bf {\hat A}}_{\tau}^\ast}
\def\Atpuv{{\bf {\hat A}}_{{\tau}^{\prime}}}
\def\Atpsuv{{\bf {\hat A}}_{{\tau}^{\prime}}^\ast}
\def\ct{c_\tau}

\def\ctt{\cos {\theta}_{\tau}}
\def\ttt{\tan {\theta}_{\tau}}
\def\stt{\sin {\theta}_{\tau}}
\def\ttp{\tan {\theta}_{{\tau}^{\prime}}}
\def\stt{\sin {\theta}_{\tau}}
\def\ctp{\cos {\theta}_{{\tau}^{\prime}}}
\def\stp{\sin {\theta}_{{\tau}^{\prime}}}
\def\at{{\alpha}_{\tau}}
\def\bt{{\beta}_{\tau}}
\def\atp{{\alpha}_{{\tau}^{\prime}}}
\def\btp{{\beta}_{{\tau}^{\prime}}}
\def\tj{ $t$-$J$ }
\def\so{{\sigma}^0}
\def\vpt{{\varphi}_{\tau}}
\def\atu{{\bf {\hat a}}_{\tau}}
\def\Au{{\bf {\hat A}}}
\def\Bu{{\bf {\hat B}}}
\def\au{{\bf {\hat a}}}
\def\bu{{\bf {\hat b}}}
\def\cu{{\bf {\hat c}}}
\def\du{{\bf {\hat d}}}
\def\tt{{\theta}_{\tau}}
\def\x{{\theta}_{{\tau}^{\prime}}}
\def\aa{a_{\tau}}
\def\aap{a_{{\tau}^{\prime}}}
\def\Ao{A_0}

\def\Bos{B_0^{\ast}}
\def\Co{C_0}
\def\Cos{C_0^{\ast}}
\def\Do{D_0}
\def\Dos{D_0^{\ast}}
\def\Bus{{\Bu}^{\ast}}
\def\Cu{{\bf {\hat C}}}
\def\Du{{\bf {\hat D}}}
\def\Cus{{\Cu}^{\ast}}
\def\Dus{{\Du}^{\ast}}
\def\St{{1\over {\sqrt 2}}}
\def\tp{t^{\prime}}
\def\ca{\cos {\theta}_1}
\def\cb{\cos {\theta}_2}
\def\sa{\sin {\theta}_1}
\def\sb{\sin {\theta}_2}
\def\caa{\cos {\phi}_1}
\def\cbb{\cos {\phi}_2}
\def\saa{\sin {\phi}_1}
\def\sbb{\sin {\phi}_2}
\def\tp{t^{\prime}}
\def\Aus{{\bf {\hat A}}^{\ast}}
\def\Aup{{\bf {\hat A}}^{\prime}}
\def\Aups{{\bf {\hat A}}^{\prime\ast}}
\def\ap{a^{\prime}}
\def\cta{c_{{\tau}^{\prime}}}
\def\ta{{\tau}^{\prime}}
\def\sib{{\bar \sigma}}


\centerline{\bf\vbig Mapping spin-charge separation without constraints}

\vskip 0.3in

\centerline{\bbig J.I. Chandler and J.M.F. Gunn}
\vskip 0.2in
\centerline{\it University of Birmingham, School of physics and space research}
\centerline{\it Edgbaston, Birmingham B15 2TT, UK}

\vskip 0.25in

\centerline{\bf\bbig Abstract}

{\med The general form of a mapping of the spin and charge degrees 
of freedom of electrons onto spinless fermions and local `spin'-${1\over 2}$ 
operators is derived. The electron Hilbert space is mapped onto a tensor product  spin-charge Hilbert space. The single occupancy condition of the \tj model is
satisfied exactly without the constraints between the operators required with 
slave particle methods and the size of the Hilbert space (four states per site) 
is conserved. The connection and 
distinction between the physical electron spin and the ``pseudospin'' used in 
these maps is made explicit. Specifically the pseudospin generates rotations 
{\it both} 
in spin space 
and particle-hole space. A geometric description (up to sign) is provided using
two component spinors. The form of the mapped $t$-$J$ Hamiltonian involves the  coupling of spin and spinless fermion currents, as one expects.}

\vskip 0.25in

\centerline{\bf\bbbig Section 1 Introduction}

\vskip 0.1in

\par

The 2D Hubbard model and the \tj model are two of the most intensely studied 
models in condensed matter physics. It has been argued that 
these models provide the minimal description of the $\rm {CuO_2}$ planes common 
to all the cuprate superconductors [1]. In one dimension the Hubbard model can 
be solved exactly and the ground state is not a Fermi liquid but shows 
separation of spin and charge degrees of freedom [2]. It has been suggested [3] 
that this spin-charge separation may also occur in two dimensions and is 
responsible for the unusual normal state properties found in the cuprate 
superconductors.

The size of the magnetic moments in the undoped cuprates implies
that the Hubbard model should be regarded as in the strong coupling limit.
In that case, to order $t^2/U$,  we may use the \tj model. 
We write the \tj model as follows
$$  {\cal H}_{t-J}=-t\sum_{\la ij \ra \si} (1-n_{i\sib})
c_{i\si}^{\dg} c_{j\si} (1-n_{j\sib}) + \hbox{ H. c. } + 
J\sum_{\la ij \ra}
\left( {\bf S}_i \cdot {\bf S}_j - {1\over 4} n_i n_j \right) \eqno (1) $$
where $\sib = -\si$, $c_{i\si}^{\dg}$ ($c_{i\si}$) are the electron creation 
(annihilation) operators, \hfill\break ${\bf S}_i = {1\over 2} \sum_{\si\sib} 
c_{i\si}^{\dg} {\si}_{\si\sib} c_{i\sib} $ are the electron spin operators and 
$n_i = \sum_{\si} c_{i\si}^{\dg} c_{i\si}$ are the total number operators for 
site $i$.

The use of constrained electron operators shows explicitly that the 
action of the model is restricted to the singly occupied sector of Hilbert
space, $ \sum_{\si} c_{i\si}^{\dg} c_{i\si} \leq 1$. (The inequality implies 
this
is a non-holonomic 
constraint.) The first techniques used to handle the occupation 
constraint in an appealing way were the slave particle methods [4,5]. The basis 
of these methods is to factorise the electron operator in terms of separate spin 
and charge operators, then the allowed states are created from a new fictitious 
vacuum. In the case of the slave boson representation [4] we have the following 
mapping of the electron operators
$$ c_{\si}=e^{\dg} f_{\si} + \si f_{\sib}^{\dg} d \eqno (2) $$
where $e^{\dg}$ and $d^{\dg}$ are bosonic operators creating empty and doubly 
occupied states respectively and $f_{\si}^{\dg}$ are fermionic operators 
creating singly occupied states. For this to satisfy the fermionic
anticommutation relations the following constraint must be satisfied
$$ e^{\dg}e + \sum_{\si} f_{\si}^{\dg} f_{\si} + d^{\dg}d =1 \eqno (3) $$
on every lattice site. In the slave fermion case [4] we have
$$ c_{\si}=f^{\dg}a_{\si} + \si a_{\sib}^{\dg} h \eqno (4) $$
Where $f^{\dg}$ and $h^{\dg}$ are spinless fermion operators creating empty
and doubly occupied states respectively and $a_{\si}^{\dg}$ are bosonic 
operators creating singly occupied states. In this case the constraint
$$ f^{\dg}f + \sum_{\si} a_{\si}^{\dg} a_{\si} + h^{\dg}h = 1 \eqno (5) $$
must be obeyed at every site. When the on-site coulomb repulsion is large
($ U\rightarrow\infty$) the operators creating doubly occupied sites drop out
leaving more elementary constraint equations.

In this way the difficult non-holonomic constraints are replaced by more 
straight forward holonomic constraints and the separation of spin and charge 
degrees of freedom can studied readily at the mean field level. 
However, these methods replace the local
constraints by an average global constraint. Then the mapped electron 
operators of equations (2) or (4) no longer satisfy the 
fermionic anticommutation relations and so are no longer a correct 
representation of electrons. This casts doubt on the results of any mean field 
calculations using slave particle methods.

Another approach is to represent the electron operators in terms of spinless
fermions and Pauli spin-${1\over 2}$ operators (or equivalently hard core 
bosons) [6,7,8].
The motivation behind this is that at half filling the \tj model reduces
to a 2D antiferromagnetic Heisenberg model for which a first order spin wave 
expansion about the classical groundstate provides an acceptable description[9].
It is hoped that 
away from half filling mean field theories based on spin wave
expansions will continue to be useful. In addition, there is the 
advantage that the single occupancy constraint is automatically satisfied 
without the need for constraint equations. In this approach the Hilbert space of
the original lattice electrons is mapped onto a tensor product hole-spin space,
Richard and Yushankha{\"{\i}}[10], for example, use the correspondence
$$ \left|\ua\ra\right. \rightarrow \left|0\ua\ra\right. \quad , \quad
\left|\da\ra\right. \rightarrow \left|0\da\ra\right. \quad , \quad
 \left|0\ra\right. \rightarrow \left|1\ua\ra\right. \eqno (6) $$
for the restricted Hilbert space of the \tj model.

The size of the Hilbert space 
is unchanged, but now we must associate a spin with every site, even unoccupied 
sites. They then use the following mapping of the constrained electron operators
$$ {\tilde c}_{\ua} = \fd (\textstyle{1\over 2} + s^z ) \quad , \quad
{\tilde c}_{\da} = \fd s^+ \eqno (7) $$
where $ {\tilde c}_{\sigma} = (1-n_{\sib})c_{\sigma} $ and the relationship 
between the true electron spin operators ${\bf S}_i$ and the ``pseudospin'' 
operators $ {\bf s}_i$ is ${\bf S}_i = (1-n_i){\bf s}_i$, where $n_i = {\fd}_i 
f_i$ is the spinless fermion number operator. So the pseudospin ${\bf s}_i$
plays a dual role, representing the real spin on sites where $n_i =1$ and 
determining whether a site is empty or doubly occupied on sites where $n_i =0$.

Using this mapping the \tj model is expressed in terms of spinless fermions and 
local spin operators with no constraints between the two. As discussed by Wang 
and Rice [11] the \tj model under this mapping lacks the time reversal symmetry 
of the original and yet they derive a model which has time reversal. Their model 
is 
also invariant under global rotations in pseudospin space ( $[ {\bf s}, {\cal 
H}_{t-J}]=0$) which the original \tj Hamiltonian is not, as will be discussed 
later.

Our work generalises mappings of this form. We derive the most general
bilinear maps from 
electron operators (not merely just the constrained electron operators) onto 
spinless fermions and local spin operators and provide an interpretation of 
these maps.

The paper is organised as follows. In section 2 we begin with a general form of 
the mapping containing undetermined scalars and vectors which we then demand 
satisfies the exact fermionic anticommutation relations. This provides us with a 
set of equations for the unknown scalars and vectors which we then 
solve for a special case. In section 3 we interpret these solutions and connect 
them
to results of other workers, giving an example in section 4. In section 5 
we show that no other types of solution are permitted and we conclude in section 
6.

\vskip 0.25in

\centerline{\bf\bbbig Section 2 Main Calculation}

\vskip 0.1in

In this section we carry out the main calculation in order to find the
generally allowed form of a mapping of spin-$1\over 2$ fermions onto spinless
fermions and the spin-$1\over 2$ Pauli spin matrices.

We begin with the following representation for the electron annihilation
operator
$$ \ct = \Pt f + \Qtd \fd. \eqno (8) $$
Where $\{f,\fd\}=1$, $\tau$ is a spin label and $\Pt$ and $\Qtd$ are operators
determined by insisting that $\ct$ obeys the standard spin-$1\over 2$ fermionic 
anticommutation relations. We begin with this form because earlier work has
ruled out any simpler mapping, for instance merely using the $f$ term above.

From the anticommutator  $\{ \ct , \cta \} = 0 $ it follows that
$$ \{ \Pt , \Qtpd \} + \{ \Ptp , \Qtd \} = 0 \eqno (9) $$ and
$$ [\Pt , \Qtpd ] + [\Ptp , \Qtd ] = 0 . \eqno (10) $$
In obtaining the above results we have used 
$f\fd = {{1}\over{2}}\{ f,\fd \} + {{1}\over{2}}[ f,\fd ]$, $\fd f = 
{{1}\over{2}}\{ f,\fd \} - {{1}\over{2}}[ f,\fd ]$ and have 
equated operator and c-number parts separately.

The anticommutator $ \{ \ct , {\cta}^{\dg} \} = {\dl}_{\tau , \ta } $ provides 
the conditions
$$ {1\over 2} \{ \Pt , \Ptpd \} + {1\over 2} \{ \Qtd , \Qtp \} =
{\dl}_{\tau\ta} \eqno (11) $$ and
$$ [\Pt , \Ptpd ] + [ \Qtp , \Qtd ] = 0. \eqno(12) $$
Equations (9),(10),(11) and (12) are the starting point of our investigation
into the allowed form of the general mapping given in equation (8).

We begin with the following forms for $\Pt$ and $\Qtd$
$$ \Pt = \Aot \so + \Atv \cdot \bsigma \eqno(13) $$ and
$$ \Qtd = \Bots \so + \Btsv \cdot \bsigma \eqno (14) $$
where $\so$ is the $2 \tm 2$ unit matrix, $\bsigma = {\si}^x {\bf {\hat x}} +
{\si}^y {\bf {\hat y}} + {\si}^z {\bf {\hat z}}$, $\Aot$ and $\Bots$ are 
c-numbers and
$\Atv$ and $\Btsv$ are complex vectors.
Again earlier work has shown that a simpler form is not possible. Equations (13) 
and (14) look like the inner product between two $4$-vectors and indeed this 
observation can be used to interpret the mapping as will be discussed later.

Initially we examine the case where $\tau = \ta$, using equation (10) and the 
identity \hfill\break
${\si}^{i} {\si}^{j} = {\dl}^{ij} + i{\ep}^{ijk} {\si}^{k}$ then leads to the 
result
$$ \Atv {\bf \tm} \Btsv = 0. \eqno (15) $$

This equation can be satisfied by letting $\Atv = \At\Atuv $ and $\Btsv = 
\Bts\Atuv$ where \hfill\break
$ \Atuvs \cdot \Atuv = 1 $.

Equation (12) with $\tau = \ta $ now leads to
$$ \left( {|\At |}^2 - {|\Bt |}^2 \right) \Atuv \tm \Atuvs = 0. \eqno (16) $$
So either $ {|\At |}^2 = {|\Bt |}^2 $ or $ \Atuv = e^{i\vpt} \atu $ where $\atu$
is a real unit vector.
Next equation (9) provides the conditions
$$ \Aot\Bots + \At\Bts\Atuv\cdot\Atuv = 0 \eqno (17) $$ and
$$ \Aot\Bts + \Bot\At = 0. \eqno (18) $$
The final constraint equation (11) leads to
$$ {|\Aot |}^2 + {|\At |}^2 + {|\Bot |}^2 + {|\Bt |}^2 = 1 \eqno (19) $$ and
$$ (\Aot\Ats + \Bots\Bt ) \Atuvs + (\Aots\At + \Bot\Bts ) \Atuv = 0.
\eqno (20) $$ 
We can solve these constraint equations for two interesting special cases and
we will discuss these solutions now before moving on to consider other possible
solutions.

If we allow $\Atuv$ to be a null vector ($ \Atuv\cdot\Atuv = 0 $) then equation 
(16) must be solved by demanding that $ {|\At |}^2 = {|\Bt |}^2 $, in which case 
we have $ \At = a_{\tau} e^{i\at}$ and $ \Bts= a_{\tau} e^{-i\bt}.$

Equation (20) then leads to $\Aot = \Bot = 0$ and equation (19)
tells us that $ a_{\tau} = {1\over {\sqrt 2}} $. This gives the following form
for the electron operator
$$ \ct = {1\over {\sqrt 2}} \left( e^{i\at} \Atuv\cdot\bsigma f + e^{-i\bt}
\Atuv\cdot\bsigma\fd \right) . \eqno (21) $$
The other special case occurs when $\Atuv$ is some multiple of a real unit 
vector \hfill\break
$\Atuv = \atu e^{i\vpt}$, then $\Atuv\cdot\Atuv = e^{2i\vpt}$ and
equation (17) says that $ \Aot\Bots + \At\Bts e^{2i\vpt} = 0$ and on using 
equation (18) we obtain $\At =\pm e^{-i\vpt}\Aot$ and 
$\Bts =\mp e^{-i\vpt}\Bots$. Equations (19) and (20) can then be used to obtain
$ {|\Aot |}^2 = {|\Bot |}^2 = a^2 = {1\over 4}. $ Putting these results
together we find the following form for the electron operator
$$ \ct = {1\over 2} \left( e^{i\at} ( 1\pm \atu\cdot\bsigma ) f +
e^{-i\bt} (1\mp \atu\cdot\bsigma ) \fd \right). \eqno (22) $$
Now we will go on to examine the $ \tau \ne {\tau}^{\prime} $ constraints for 
these special cases to see if they are valid representations of the electron 
operators. The cases where $\ct$ and $c_{{\tau}^{\prime}}$ are both represented 
by either real or null vectors may be readily shown to be inconsistent. Thus we 
allow $ \ct $ to be represented using a null vector with
$ \Pt = {1\over {\sqrt 2}} e^{i\at} \Au\cdot\bsigma $ and
$ \Qtd = {1\over {\sqrt 2}} e^{-i\bt} \Au\cdot\bsigma$ and 
$ c_{{\tau}^{\prime}}^{\dg} $ to be represented using a real vector with 
$ \Ptpd = {1\over 2} e^{i\ga} (1\pm\Bu\cdot\bsigma ) $ and $ \Qtpd = {1\over 2} 
e^{-i\dl} (1\mp\Bu\cdot\bsigma ) $. Then equation (10) leads to 
$$ \left( e^{i(\al - \dl )} + e^{i(\ga - \be )} \right) \Au\tm\Bu = 0. \eqno 
(23) $$ 
But since $\Au$ is a null vector and $\Bu$ is some multiple of a real vector
we must satisfy this condition non-trivially ($\Au\not= 0$ and/or $\Bu\not= 0$) 
by demanding that
$$ e^{i(\al - \dl )} + e^{i(\ga - \be )} = 0. \eqno (24) $$
Next we use equation (9) to obtain the constraint
$$ \left( e^{i(\al - \dl )} - e^{i(\ga - \be )} \right) \Au\cdot\Bu = 0
\eqno (25) $$ and
to avoid inconsistencies we must demand that
$$ \Au\cdot\Bu = 0. \eqno (26) $$
Equation (12) gives the condition
$$ \left( e^{i(\al - \ga )} + e^{i(\dl - \be )} \right) \Au\tm\Bu = 0. 
\eqno (27) $$
Now we cannot have $\Au\tm\Bu =0$ except in a trivial case therefore we are 
lead to the constraint
$$ e^{i(\al - \ga )} + e^{i(\dl - \be )} = 0 \eqno (28) $$
which is identical to equation (24).
Equation (11) can be satisfied by equation (24) and equation (26) and so does 
not lead to any new constraints.
We can satisfy equation (15), ensure that $\Bu$ is a real vector and that $\Au$ 
is null by letting
$$ \Au = {1\over {\sqrt 2}} (\au + i\bu ) \quad ; \quad \au\cdot\bu = 0 \eqno 
(29) $$
$$ \Bu = \cu \quad ; \quad \cu = \pm\au\tm\bu . \eqno (30) $$
Where $\au$ , $\bu$ and $\cu$ are real unit vectors (also from the above 
$\Au\cdot {\Au}^\ast =1 $). Equation (24) is solved by taking logarithms 
resulting in
$$ (\al + \be ) = (\ga + \dl \pm \pi ) \eqno (31) $$
which we satisfy by setting
$$ \al = \tt + \phi \quad ; \quad \be = -\tt + \phi \pm \pi \eqno(32) $$
$$ \ga = \x + \phi \quad ; \quad \dl = -\x + \phi . $$

Putting all these results together we arrive at the final forms of the
representations of the electron operators
$$ \eqalign{\ct^{\rm null} &= {{e^{i\tt}}\over 2} (\au + i\bu )\cdot\bsigma 
\left(
e^{i\phi} f - e^{-i\phi} \fd \right) \cr
 c_{\tau'}^{\rm real} &= {{e^{i\x}}\over 2} \left( e^{i\phi} 
(1\pm\cu\cdot\bsigma ) f + e^{-i\phi} ( 1\mp\cu\cdot\bsigma ) \fd \right)\cr} 
\eqno(33) $$
or 
$$ \eqalign{c_{\tau'}^{\rm null} &= {{e^{i\x}}\over 2} (\au +i\bu )\cdot\bsigma
\left( e^{i\phi} f + e^{-i\phi} \fd \right) \cr
 \ct^{\rm real} &= {{e^{i\tt}}\over 2} \left( e^{i\phi} (1\pm\cu\cdot\bsigma 
) f
- e^{-i\phi} (1\mp\cu\cdot\bsigma ) \fd \right) .\cr} \eqno (34) $$
These are the basic results for this special case. In the next section we
will interpret these results and show that they can be described in terms of
the vectors which are used in the geometric description (up to sign) of spinors.
In section 4 we show that there are no further allowed solutions.

\vskip 0.25in

\centerline{\bf\bbbig Section 3 Interpretation in terms of spinors}

\vskip 0.1in

We can understand the form of these mappings by making use of the relationship
between spinors  ($X^{a{\dot u}}$) with one dotted and one undotted index and 
$4$-vectors [12], where an undotted index refers to transformation by a Lorentz 
spin transform matrix (SL(2,C)) and the dotted index refers to transformation by 
a complex conjugated Lorentz spin transform matrix.
This can be done because of the group homomorphism between SL(2,C) matrices
and matrices ${\cal L_{\ua}^+}$ representing proper orthochronous Lorentz 
transformations.

The explicit connection between a $4$-vector $x^{\mu}$ and a two component 
spinor $X^{a{\dot u}}$ is
the following
$$ X^{a{\dot u}} = x^{\mu} {\si}_{\mu}^{a{\dot u}} \quad ; \quad x^{\nu} = 
-{1\over 2} {\sigma}_{a{\dot u}}^{\nu} X^{a{\dot u}} \eqno (35) $$
where $ {\sigma}_{\mu} = ( {\sigma}_0 , \bsigma ). $ 

Given any two component spinor ${\xi}^a$, we can define its spinor mate 
${\eta}^a$ by ${\xi}_a {\eta}^a = 1$ where ${\xi}_a = {\xi}^b {\ep}_{ba}$ and
$$ {\ep}_{ba} = \left( \matrix{ 0 & 1 \cr -1 & 0 \cr }\right). \eqno (36) $$
Clearly the choice of spinor mate is not unique and in general 
$ {\eta}_{\hbox{\small new}}^a = {\eta}^a + {\al} {\xi}^a$. \hfill\break
We can then generate from these two spinors, $4$-vectors which
are useful in the geometrical description (up to a sign) of one component 
spinors.
 
Firstly we can generate the ``null flagpole vector'' $x^{\mu}$ defined via
$$ X^{a{\dot u}} = {\xi}^a {{\bar \xi}}^{{\dot u}} = x^{\mu} {\si}_{\mu}^{a{\dot 
u}} \eqno (37) $$
where the overbar denotes complex conjugation. Here $x^{\mu}$ is a future 
directed null vector ( $x^{\mu} x_{\mu} = 0$ )
and so defines a 3D hypercone or lightcone. We can define an analogous
vector $w^{\mu}$ using the spinor mate ${\eta}^a$ via
$$ W^{a{\dot u}} = {\eta}^a {{\bar \eta}}^{{\dot u}} = w^{\mu} 
{\si}_{\mu}^{a{\dot u}}. \eqno (38) $$
Next we define the spacelike ``flag'' vector $y^{\mu}$ via
$$ Y^{a{\dot u}} = {\xi}^a {{\bar \eta}}^{{\dot u}} + {\eta}^a {{\bar 
\xi}}^{{\dot u}} 
= y^{\mu} {\si}_{\mu}^{a{{\dot u}}}. \eqno (39) $$
This vector is orthogonal to the flagpole vector ($y^{\mu} x_{\mu} = 0$).
Because the choice of spinor mate ${\eta}^a$ was not unique, $y^{\mu}$ is not
unique and we can have in general \hfill\break
$ y_{\hbox{\small new}}^{\mu} = y^{\mu} + ( {\al} + {{\bar \alpha}} ) x^{\mu}$. 
So the possible flag vectors $y^{\mu}$ are all coplanar, and orthogonal to 
$x^{\mu}$ the flagpole vector.

Finally we generate the spacelike $4$-vector $z^{\mu}$ via
$$ Z^{a{\dot u}} = i {\xi}^a {{\bar \eta}}^{{\dot u}} -i {\eta}^a {{\bar 
\xi}}^{{\dot u}}
= z^{\mu} {\si}_{\mu}^{a{{\dot u}}}. \eqno (40) $$
This $4$-vector is orthogonal to both $x^{\mu}$ and $y^{\mu}$, so $y^{\mu}$
and $z^{\mu}$ are basis vectors in the 2D space on the lightcone
orthogonal to $x^{\mu}$.

The general spinor of fixed magnitude ${\xi}^a$ is obtained by rotating the 
familiar spinor $\left|\ua\ra\right. = \left( \matrix{ 1 \cr 0 \cr }\right)$
which has flagpole vector $ x^{\mu} = (s,s {\bf {\hat z}}) $  in the positive 
sense  through the three Euler angles $\theta$,$\phi$ and $\psi$ [13] resulting 
in 
$$ {\xi}^a = \sqrt{ 2s } \left( \matrix{ \cos {{\theta}\over 2} 
e^{i({{\phi + \psi }\over 2})} \cr \sin {{\theta}\over 2}
e^{i({{\phi - \psi }\over 2})} \cr }\right). \eqno (41) $$
The spinor mate is given by
$$ {\eta}^a = \sqrt{ 2s } \left( \matrix{ -\sin {{\theta}\over 2}
e^{i({{\psi - \phi}\over 2})} \cr \cos {{\theta}\over 2}
e^{i({{\phi - \psi}\over 2})} \cr }\right). \eqno (42) $$

Using equation (37) we find that the flagpole vector is given by
$$ x^{\mu} = s \left( \matrix{ 1 \cr \sin\theta\cos\phi \cr \sin\theta\sin\phi
\cr \cos\theta \cr }\right) \eqno (43) $$                                        
                                        as would be expected.
Next we use equation (38) to find
$$ w^{\mu} = s \left( \matrix{ 1 \cr -\sin\theta\cos\phi \cr -\sin\theta\sin\phi
\cr -\cos\theta \cr }\right) . \eqno (44) $$
Equation (39) leads to
$$ y^{\mu} = 2s \left( \matrix{ 0 \cr \cos\theta\cos\phi\cos\psi - 
\sin\phi\sin\psi
\cr \cos\theta\sin\phi\cos\psi + \cos\phi\sin\psi \cr -\sin\theta\cos\psi \cr
}\right) \eqno (45) $$
and lastly equation (40) gives
$$ z^{\mu} = 2s \left( \matrix{ 0 \cr \cos\theta\cos\phi\sin\psi + 
\sin\phi\cos\psi \cr \cos\theta\sin\phi\sin\psi - \cos\phi\cos\psi \cr
-\sin\theta\sin\psi \cr }\right). \eqno (46) $$

We can use the above analysis to describe the form of the mappings given in
equations (33) and (34). 

In that case the $4$-vector $( s,s\cu )$ corresponds to the flagpole vector
$x^{\mu}$ of the spinor we are representing while the $4$-vector $ (s, -s\cu )$
is the flagpole vector of the spinor mate $w^{\mu}$. The $4$-vectors
$2s ( 0,\au )$ and $2s ( 0,\bu )$ then correspond to the flag vectors $y^{\mu}$
and $z^{\mu}$ respectively. This means that we can rewrite the mapped electron
operators of equation (34) in the following ways
$$ c_{\xi} = e^{i{{\theta}_{\xi}}} ( {\eta}^a {{\bar \xi}}^{{\dot u}} )
( e^{i\phi} f + e^{-i\phi} \fd ) $$ 
$$ c_{\eta} = e^{i{{\theta}_{\eta}}} \left( ( {\eta}^a {{\bar \eta}}^{{\dot u}} 
)
e^{i\phi} f - ( {\xi}^a {{\bar \xi}}^{{\dot u}} ) e^{-i\phi} \fd \right)
\eqno (47) $$
or alternatively
$$ c_{\xi} = {1\over 2} e^{i{{\theta}_{\xi}}} ( y^{\mu} + i z^{\mu} ) 
{\sigma}_{\mu}
( e^{i\phi} f + e^{-i\phi} \fd ) $$
$$ c_{\eta} = e^{i{{\theta}_{\eta}}} ( w^{\mu} {\sigma}_{\mu} e^{i\phi} f  - 
x^{\mu} {\sigma}_{\mu} e^{-i\phi} \fd ) \eqno (48) $$
where a particular choice of sign has been made in equation (34).
The two important points to notice about the above equations are that the 
$4$-vectors which emerge are the natural ones used in the geometrical
description of the spinor we are attempting to represent, and that only the 
spinor and its spinor mate are required to find the appropriate form of the 
mapping.

\vskip 0.25in

\centerline{\bf\bbbig Section 4 Example}

We now take a specific example to illustrate the above ideas more clearly.
Consider the standard spinor basis for which ${\sigma}_z$ is diagonal, that is
$$ {\xi}^a = \left( \matrix{ 1 \cr 0 \cr }\right) = \left|\ua\ra\right. \quad ; 
\quad  {\eta}^a = \left( \matrix{ 0 \cr 1 \cr }\right) = \left|\da\ra\right. . 
\eqno (49) $$
The complete spin charge direct product basis in this case is as follows
$$ |0\da\ra = \left|\da\ra\right. \otimes |0\ra = \left( \matrix{ 0 \cr 1 \cr } 
\right)
\otimes \left( \matrix{ 0 \cr 1 \cr } \right) = \left( \matrix{ 0 \cr 0 \cr 0 
\cr 1 \cr } \right) \eqno (50) $$
$$ |0\ua\ra = \left|\ua\ra\right. \otimes |0\ra = \left( \matrix{ 1 \cr 0 \cr } 
\right)
\otimes \left( \matrix{ 0 \cr 1 \cr } \right) = \left( \matrix{ 0 \cr 1 \cr 0
\cr 0 \cr } \right) \eqno (51) $$
$$ |1\da\ra = \left|\da\ra\right. \otimes |1\ra = \left( \matrix{ 0 \cr 1 \cr } 
\right)
\otimes \left( \matrix{ 1 \cr 0 \cr } \right) = \left( \matrix{ 0 \cr 0 \cr 1
\cr 0 \cr } \right) \eqno (52) $$ and
$$ |1\ua\ra = \left|\ua\ra\right. \otimes |1\ra = \left( \matrix{ 1 \cr 0 \cr } 
\right)
\otimes \left( \matrix{ 1 \cr 0 \cr } \right) =\left( \matrix{ 1 \cr 0 \cr 0
\cr 0 \cr } \right) \eqno (53) $$
and we can make the following identification between the
original states and the new direct product states given above
$$ \left|0\ra\right. \rightarrow \left|0\da\ra\right. \quad , \quad 
\left|\ua\da\ra\right. \rightarrow \left|0\ua\ra\right. \quad , \quad 
\left|\da\ra\right. \rightarrow \left|1\da\ra\right. \quad , \quad 
\left|\ua\ra\right. \rightarrow \left|1\ua\ra\right. \quad\eqno(54) $$
Then it follows from equations (37)-(40) that 
$x^{\mu} = (s,s{\bf {\hat z}})$, $w^{\mu} = (s,-s{\bf {\hat z}})$, 
$y^{\mu} = (2s,2s{\bf {\hat x}})$ and $z^{\mu} = (2s,-2s{\bf {\hat y}}).$
Now if we let $ {\theta}_{\xi} = {\theta}_{\eta} = \phi = 0 $ in equation (48) 
we obtain the following mapping of the electron operators
$$ c_{\ua} = {1\over 2}{\si}^- (f+\fd) \eqno (55) $$ and
$$ c_{\da} = {1\over 2} (1-{\si}^z ) f - {1\over 2} (1+{\si}^z ) \fd . 
\eqno (56) $$
This representation has the correct behaviour when acting on the spin charge 
direct product states as may easily be checked.\hfill\break
The constrained electron operators  
$ {\tilde c}_{\sigma} = (1-n_{-\sigma})c_{\sigma} $ are given by
$$ {\tilde c}_{\ua} = {1\over 2}f{\si}^- \quad , \quad {\tilde c}_{\da} = {1\over 
2}
f(1-{\si}^z) \eqno (57) $$ 
Which is the analogue of equation (7) for the choice of pseudospin on the empty 
site we have made.

The inverse of this mapping may also be obtained, starting from the fact that
\hfill\break $c_{\da}^{\dg}-c_{\da} = \fd - f$ and $(c_{\da}^{\dg} + 
c_{\da})(1-2c_{\ua}^{\dg}c_{\ua}) = \fd + f $ we can obtain
$$ \fd = c_{\da}^{\dg}(1-c_{\ua}^{\dg}c_{\ua})-c_{\da}c_{\ua}^{\dg}c_{\ua}
\eqno (58) $$ and
$$ f = (1-c_{\ua}^{\dg}c_{\ua})c_{\da}-c_{\ua}^{\dg}c_{\ua}c_{\da}^{\dg}.
\eqno (59) $$
Starting from ${\si}^+ = 2c_{\ua}^{\dg}(f+\fd)$ we can also obtain
$$ \eqalign{{\si}^+ &= 2( c_{\ua}^{\dg}c_{\da}^{\dg}+c_{\ua}^{\dg}c_{\da}) \cr
 {\si}^- &= 2( c_{\da}c_{\ua}+c_{\da}^{\dg}c_{\ua}) \cr 
 {\si}^z &= 2c_{\ua}^{\dg}c_{\ua}-1.\cr} \eqno (60) $$
It is interesting to note that these results may be written more succinctly as
$$ \bsigma = 2({\bf S}+{\bf J}) \eqno (61) $$
where ${\bf S}$ is the true electron spin operator and ${\bf J}$ is the 
generator of rotations in particle-hole space (``isospin'' operator), whose 
explicit representation is given by 
$$ \eqalign{J^+=&J^x+iJ^y=c_{\ua}^{\dg}c_{\da}^{\dg}\cr
J^-=&J^x-iJ^y=c_{\da}c_{\ua}\cr
J^z=&\textstyle{1\over 2}\sum_{\si} c_{\si}^{\dg}c_{\si} -{1\over 2}\cr}. \eqno 
(62) $$
The $J^i$ form a spin algebra with the usual commutation relations.
So the spin operators appearing in our representation are not the same as those
in the standard electron representation but are composed of the true electron 
spin operators and operators which generate rotations in particle-hole space. 
Using equation (60) we can see that the \tj model commutes with the 
total $z$ component of the pseudospin only and so is not invariant with respect 
to general rotations in pseudospin space (invariant under both rotations in spin 
space and particle-hole space) unlike the model obtained by Wang and Rice [6] as 
mentioned earlier.

The standard number operators evaluated using the mapped electron operators are                                        
$$ n_{\ua} = {1\over 2}(1+{\si}^z) \quad ; \quad n_{\da} = {1\over 2}(1+{\si}^z)
-\fd f {\si}^z . \eqno (67) $$
It is also interesting to express the \tj model with the mapped electron
operators, this leads to
$$ {\cal H}_{t-J}=-t\sum_{\scriptstyle \langle ij \rangle } f_{i}^\dagger f_{j} 
(1+ \bsigma _{i} \cdot \bsigma _{j}-i( \bsigma _{i} \times \bsigma _{j}) \cdot 
{\bf {\hat z}} -\sigma_{i}^z-\sigma_{j}^z)+\hbox{ H. c. }+{J\over 
2}\sum_{\scriptstyle \langle ij \rangle } n_{i} n_{j} \bsigma _{i} \cdot \bsigma 
_{j}\eqno (68) $$
where $n_{i}$ are the spinless fermion occupation numbers $n_i = f_i^{\dg}f_i$.

Written in this form the coupling between doped holes and the spin background is 
shown explicitly with terms linking the kinetic energy density to a 
ferromagnetic pseudospin interaction and the fermionic current to a
pseudospin current. The importance of these interactions to the interpretation of the \tj model has been discussed in [14]. The conservation of the total 
number 
of spinless fermions $\sum_{i} f_i^{\dg} f_i$ reflects the conservation of the 
total number of singly occupied sites. This mapped $t$-$J$ model does not
have the time reversal invariance of the original as discussed by Wang and Rice 
[11]. 

\vskip 0.25in

\centerline{\bf\bbbig Section 5 The general case}

\vskip 0.1in

In this section we show that no solutions other than those obtained in section 2 
are allowed. $\Atuv$ will not in general be a null vector or a real unit vector
multiplied by a phase factor. Therefore let
$$ \Atuv\cdot\Atuv = t^2 e^{2ia}. \eqno (69) $$
Equations (17) and (18) yield
$$ \Bots = \pm t e^{ia} \Bts \quad ; \quad \Aot = \mp t e^{ia} \At .
\eqno (70) $$
We satisfy equation(16) by letting
$$ \At = b e^{i\al} \quad ; \quad \Bts = b e^{-i\be}. \eqno (71) $$
We then substitute equations (70) and (71) into equation (19) to obtain
$$ 2( 1+t^2 )b^2 = 1 \quad ; \quad b={1\over {\sqrt {2( 1+t^2 )}}}. 
\eqno (72)$$
We can write the above results in a more useful way by letting $ t=\ttt $, then 
we have  
$ b={1\over {\sqrt 2}}\ctt $ and $ bt={1\over {\sqrt 2}}\stt $ and 
$$ \At = {1\over {\sqrt 2}}\ctt e^{i\at} \quad ; \quad \Aot = \mp
{1\over {\sqrt 2}}\stt e^{i( \aa + \at )} \eqno (73) $$
$$ \Bts = {1\over {\sqrt 2}}\ctt e^{-i\bt} \quad ; \quad \Bots = \pm
{1\over {\sqrt 2}}\stt e^{i( \aa - \bt )}. \eqno (74) $$
The above forms also satisfy equation (20). Putting the above results together 
the general form for the representation of the electron operator is
$$ \ct = {{e^{i\at}}\over {\sqrt 2}}\left( \mp \stt e^{i\aa} + 
\ctt\Atuv\cdot\bsigma \right) f + {{e^{-i\bt}}\over {\sqrt 2}}\left( \pm \stt
e^{i\aa} + \ctt\Atuv\cdot\bsigma \right) \fd . \eqno (75) $$
To examine the $\tau \not= {\tau}^{\prime}$ constraints we set
$$ \Pt = \Ao + \Au\cdot\bsigma \quad ; \quad \Qtd = \Bos + \Bus\cdot\bsigma $$
$$ \Ptp = \Co + \Cu\cdot\bsigma \quad ; \quad \Qtpd = \Do + \Du\cdot\bsigma
\eqno (76) $$ where
$$ \eqalign{\Ao &= \mp\St\stt e^{i( \aa + \at )} \quad ; \quad \Au = \St\ctt 
e^{i\at}\Atuv 
\cr
\Bos &= \pm\St\stt e^{i( \aa - \bt )} \quad ; \quad \Bus = \St\ctt e^{-i\bt}
\Atuv\cr
\Co &= \mp\St\stp e^{i( \aap + \atp )} \quad ; \quad \Cu = \St\ctp
e^{i\atp} \Atpuv \cr
\Dos &= \pm\St\stp e^{i( \aap - \btp )} \quad ; \quad \Dus = \St\ctp
e^{-i\btp} \Atpuv \cr}. \eqno (77) $$
We then use equations (9)-(12) to obtain the following constraint
equations
$$ \Au\times\Dus + \Cu\times\Bus = 0 \eqno (78) $$
$$ \Au\times\Cus + \Du\times\Bus = 0 \eqno (79) $$
$$ \Au\Dus + \Dos\Au + \Co\Bus + \Bos\Cu = 0 \eqno (80) $$
$$ \Ao\Dos + \Co\Bos + \Au\cdot\Dus + \Cu\cdot\Bus = 0 \eqno (81) $$
$$ \Ao\Cus + \Cos\Au + \Dos\Bus + \Bos\Du = 0 \eqno (82) $$
$$ \Ao\Cos + \Do\Bos + \Au\cdot\Cus + \Du\cdot\Bus = 0. \eqno (83) $$
The first of these constraints becomes 
$$ {1\over 2} \ctt\ctp ( e^{i(\at - \btp )} - e^{i(\atp - \bt )} ) 
\Atuv\times\Atpuv = 0. \eqno (84) $$
We do not let $\Atuv\times\Atpuv = 0$ because this means that the two operators
for the spinor states $\tau$ and ${\tau}^{\prime}$ are essentially identical and 
is just a trivial result as is letting either $\ctt = 0$ or $\ctp =0$.
Instead we satisfy the constraint by demanding that
$$ ( \at + \bt ) = ( \atp + \btp ). \eqno (85) $$
Equations (79) and (80) are also both satisfied by equation (85) so the next
constraint comes from equation (81) and is
$$ \Atuv\cdot\Atpuv = \pm\ttt\ttp e^{i( \aa +\aap )}. \eqno (86) $$
Equation (82) is also satisfied by equation (85) and so the final constraint
comes from equation (83) and is
$$ \Atuv\cdot\Atpsuv = \mp\ttt\ttp e^{i( \aa - \aap )}. \eqno (87) $$
Letting $\ttt = t$, $\ttp = \tp$, $\at = a$ and $\atp = \ap$ to simplify the 
following work, we can group together the constraint equations as follows
$$\eqalign{ \Au\cdot\Aus = 1 \quad &; \quad \Aup\cdot\Aups = 1 \cr
 \Au\cdot\Au = t^2 e^{2ia} \quad &; \quad \Aup\cdot\Aup = {\tp}^2 e^{2i\ap}\cr
\Au\cdot\Aup = \pm t\tp e^{i(a+\ap)} \quad &; \quad \Au\cdot\Aups = \mp t
\tp e^{i(a-\ap)}\cr}. \eqno (88) $$
To investigate these constraints further we write the (in general complex) unit
vectors as
$$ \Au = \St \left( \au + i\bu \right) \quad ; \quad \Aup = \St \left(
\cu +i\du \right) . \eqno (89) $$
where $\au$, $\bu$, $\cu$ and $\du$ are real unit vectors and these forms
automatically satisfy the first line of equation (88)
 
Equating the real and imaginary parts of the second line of equation (88) we are 
lead to the fact that $\cos 2a = 0$, $\au\cdot\bu = t^2 \sin 2a$ and that
$\cos 2\ap = 0$, $\cu\cdot\du = {\tp}^2 \sin 2\ap$.
From the final line of equation (88) we have the following results
$$ \eqalign{\au\cdot\cu = \mp 2t\tp \sin a \sin\ap \quad &; \quad \bu\cdot\du = 
\mp
2t\tp \cos a \cos\ap \cr
 \bu\cdot\cu = \pm 2t\tp \cos a \sin\ap \quad &; \quad \au\cdot\du = \pm
2t\tp \sin a \sin\ap \cr}\eqno (90) $$
We start by solving for $a$ and $\ap$ for which we obtain $a=\pm {{\pi}\over 4},
\pm {{3\pi}\over 4}$ and $\ap =\pm {{\pi}\over 4}, \pm {{3\pi}\over 4}$ and
so there are $16$ cases to consider but we can actually just consider the case 
where $a=\ap = {{\pi}\over 4}$ as all of the others may be obtained by 
appropriate inversions of the $4$ unit vectors. We also let $\bu\rightarrow 
-\bu$ and $\du\rightarrow -\du$ to obtain a more symmetrical set of equations.
This leaves us with the following to solve
$$\au\cdot\bu = -t^2 \quad ; \quad \cu\cdot\du = -{\tp}^2 \eqno (91) $$
$$ \au\cdot\cu = \mp t\tp \quad ; \quad \bu\cdot\du = \mp t\tp \eqno (92) $$
$$ \bu\cdot\cu = \mp t\tp \quad ; \quad \au\cdot\du = \mp t\tp . \eqno (93) $$
Firstly we note that $ ( \au\times\bu ) \cdot ( \cu\times\du ) =0 $, so that
the plane contain g $\au$ and $\bu$ is orthogonal to the plane contain g $\cu$
and $\du$ so we may set up the four vectors as follows.
$$ \au = \left( \matrix{ \ca\cr\sa\cr 0 \cr } \right) \quad ; \quad
\bu = \left( \matrix{ \cb\cr\sb\cr 0 \cr } \right) \eqno (94) $$
$$ \cu = \left( \matrix{ \caa\cr 0 \cr \saa\cr } \right) \quad ; \quad
\du = \left( \matrix{ \cbb\cr 0 \cr \sbb\cr } \right) . \eqno (95) $$
Equation (91) then becomes
$$ \cos ( {\theta}_1 - {\theta}_2 )=-t^2 \quad ; \quad \cos ( {\phi}_1 - 
{\phi}_2 ) = -{\tp}^2 . \eqno (96) $$
Equations (92) and (93) now read
$$ \cos {\theta}_1 \cos {\phi}_1 = t\tp \quad ; \quad \cos {\theta}_2 \cos
{\phi}_2 = t\tp \eqno (97) $$ and
$$ \cos {\theta}_2 \cos {\phi}_1 =t\tp \quad ; \quad \cos {\theta}_1 \cos
{\phi}_2 = t\tp \eqno (98) $$ where we have chosen the plus sign in the above.
We can obtain the two constraints
$$ {\theta}_2 = \pm {\theta}_1 \quad ; \quad {\phi}_2 = \pm {\phi}_1 
\eqno (99) $$
We cannot have ${\theta}_2 = {\theta}_1$ and/or ${\phi}_2 = {\phi}_1$ as
then from equation (91) we see that there is no real solution for $t$ and/or 
$\tp$. So there is only one case to consider, and the unit vectors become
$$ \au = \left( \matrix{ \ca\cr\sa\cr 0 \cr } \right) \quad ; \quad 
\bu = \left( \matrix{ \ca\cr -\sa\cr 0 \cr } \right) \eqno (100) $$
$$ \cu = \left( \matrix{ \caa\cr 0 \cr \saa\cr } \right) \quad ; \quad 
\du = \left( \matrix{ \caa\cr 0 \cr -\saa\cr } \right) . \eqno (101) $$
Finally we must solve the following
$$ \cos\theta\cos\phi = t\tp \eqno (102) $$
$$ \cos 2\theta = -t^2 \eqno (103) $$ 
$$ \cos 2\phi = -{\tp}^2 \eqno (104) $$
where we have dropped the unrequired numerical subscript.

The last two equations place limits on the ranges of $\theta$ and $\phi$
which are as follows
$$ {{3\pi}\over 4} > \theta > {{\pi}\over 4} \quad ; \quad {{3\pi}\over 4} >
\phi > {{\pi}\over 4} \eqno (105) $$
Equations (102)-(104) lead to the result $ {(\cos\theta)}^2 {(\cos\phi)}^2 =
\cos{2\theta} \cos{2\phi}$ which is clearly not true, so we are able to rule out 
any solutions other than those obtained in section 2.

\vskip 0.25in

\centerline{\bf\bbbig Section 6 Summary}

We have obtained all of the allowed forms of local bilinear maps of electron 
operators onto 
spinless fermion and `spin' operators. We have shown how these results may 
be interpreted in terms of the geometrical description of spinors. An important 
result of our work is an understanding of the ``pseudospin'' operator used in 
these mappings. The pseudospin operator is composed of two operators obeying 
spin-$1\over2$ algebra acting in distinct subspaces. They are shown to be the 
true electron spin operator and an ``isospin'' operator which generates 
rotations in 2D particle hole space. 

Using the simplest allowed mapping the \tj 
model is expressed in a form in which the coupling between the doped holes and 
the magnetic background is revealed. The general Hubbard model (as against the $U=\infty$ limit) has a more complicated form involving the production and annihilation of spinless fermion pairs. Only in the $U=\infty$ limit is the number of the fermions conserved. Our treatment is thus ideally suited to the strong coupling limit. Initial mean field analysis has lead to 
reasonable results.
\vskip 0.25in

\centerline{\bf\bbbig Acknowledgments}

JIC would like to thank the EPSRC for a research 
studentship. JMFG would like to thank M W Long for many discussions and the EPSRC for support 
through grant GR/J35221.

\vskip 0.25in

\centerline{\bf\bbbig References}

\vskip 0.1in

\parindent 1em

\item{[1]}  F.C. Zhang and T.M. Rice, {\it Phys. Rev.} B {\bf 37}, 3759 (1988).
\item{[2]}  T.M. Rice in: {\it Les Houches 1991 strongly interacting fermions 
and 
high temperature superconductivity}, Ed. B. Dou\c cot and J. Zinn-Justin, 
North-Holland (1995).
\item{[3]}  P.W. Anderson, {\it Phys. Rev. Lett.} {\bf 64}, 1834 (1990).
\item{[4]}  R. Fre{\'e}sard and P. W{\"o}lfle, {\it Int. J. Mod. Phys.} {\bf 
B6}, 237 
(1992).
\item{[5]}  J.C Le Guillou and E. Ragoucy, {\it Phys. Rev.} B {\bf 52}, 2403 
(1995).
\item{[6]}  G.G. Khaliullin, {\it JETP Lett.} {\bf 52}, 389 (1990).
\item{[7]}  S. Feng, Z.B. Su and L. Yu, {\it Phys. Rev.} B {\bf 49},  2368 
(1994).
\item{[8]}  A. Angelucci, S. Sorella and D. Poilblanc, {\it Phys. Lett.} A {\bf 
198}, 
145 (1995).
\item{[9]}  D.C. Mattis, {\it The theory of magnetism.-1:Statics and dynamics}, 
Springer-Verlag (1981).
\item{[10]}  J.L. Richard and Yushankha{\"{\i}}, {\it Phys. Rev.} B {\bf 47}, 
1103 
(1993).
\item{[11]}Y.R. Wang and M.J. Rice, {\it Phys. Rev.} B {\bf 49}, 4360 
(1994).
\item{[12]}  G.L. Naber, {\it The geometry of Minkowski space time}, 
Springer-Verlag (1992).
\item{[13]}C.W. Misner, K.S. Thorn and J.A. Wheeler, {\it Gravitation}, 
W.H.Freeman (1973).
\item{[14]} G. Baskaran, {\it Progress of Theoretical physics Supplement} 
No. 107, 49 (1992).
\end